\documentclass[prl,aps,twocolumn,showpacs]{revtex4}
\usepackage[dvips]{graphicx}
\usepackage{graphics}
\begin{document}

\title{Fluid Invasion in Porous Media: Viscous
Gradient Percolation}

\author{Chi-Hang Lam} 

\affiliation{
Department of Applied Physics, Hong Kong Polytechnic University, Hung
Hom, Hong Kong }
\date{\today}

\begin{abstract}
We suggest that the dynamics of stable viscous invasion fronts in porous
media depends on the volume capacitance of the media. At high volume
capacitance, our network simulations provide numerical
evidence of a scaling relation between the front width and its
velocity. In the low volume capacitance regime, we derive a new
effective scaling supported by network simulations and is in agreement
with previous experiments on imbibition in paper and collections of
glass beads.

\end{abstract}

% 47.55.Mh Flows through porous media
% 05.40.-a Fluctuation phenomena, random processes, 
% 	noise and brownian motion
% 68.35.Fx Diffusion; interface formation
% 64.60.Ak Renormalization-group, fractal, and percolation studies 
%	of phase transitions
% 05.70.Ln Nonequilibrium and irreversible thermodynamics.

\maketitle

%%%%%%%%%%%%%%%%%%%%%%%%%%%

Percolation theory has contributed much to our understanding
of immiscible fluid displacement in porous media
\cite{Sahimi,Stauffer}. For slow flow,
randomness in capillary forces dominates and invasion percolation
applies. In the presence of gravity, the local percolation probability
is spatially non-uniform and gradient percolation is the standard
description \cite{Hulin}. Alternatively, when a viscous fluid
displaces a non-viscous one, the pressure gradient also induces a
gradient in the local percolation probability. The invasion front is
hence expected to follow percolation geometry as well. However, the
pressure field is not known {\it apriori}. This problem is therefore
much more challenging and not well-understood.

Applying percolation theory, Xu, Yortsos and Salin suggested that the width
of a viscous invasion front $w$ depends on its propagation velocity $v$
according to
\begin{equation}
\label{wv}
w \sim v^{-\kappa}
\end{equation}
where $\kappa \simeq 0.38$ in two dimensions \cite{Xu}. This result is
identical to that from a Buckley-Leverett type theory of Wilkinson
\cite{Wilkinson} although there have been other suggestions
\cite{Lenormand,Blunt}. Previous network simulations lead to
inconclusive results due to the limited network sizes used
\cite{Xu,Aker}. In this paper, we report simulations at much larger
scales using the simplest possible network models. Surprisingly, we
observe two distinct scalings depending on the volume capacitance of
the network defined as the volume of liquid that can be extracted from
the porous medium locally per unit decrease in pressure
\cite{Furuberg}. At high capacitance, our simulations support the
theory in Refs. \cite{Xu,Wilkinson}. At low capacitance, we obtain a
different scaling in agreement with two interesting experiments on
imbibition in paper \cite{Horvath} and collections of glass beads
\cite{Rubio}.

% QB1
\begin{figure}
\includegraphics[width=0.49\columnwidth]{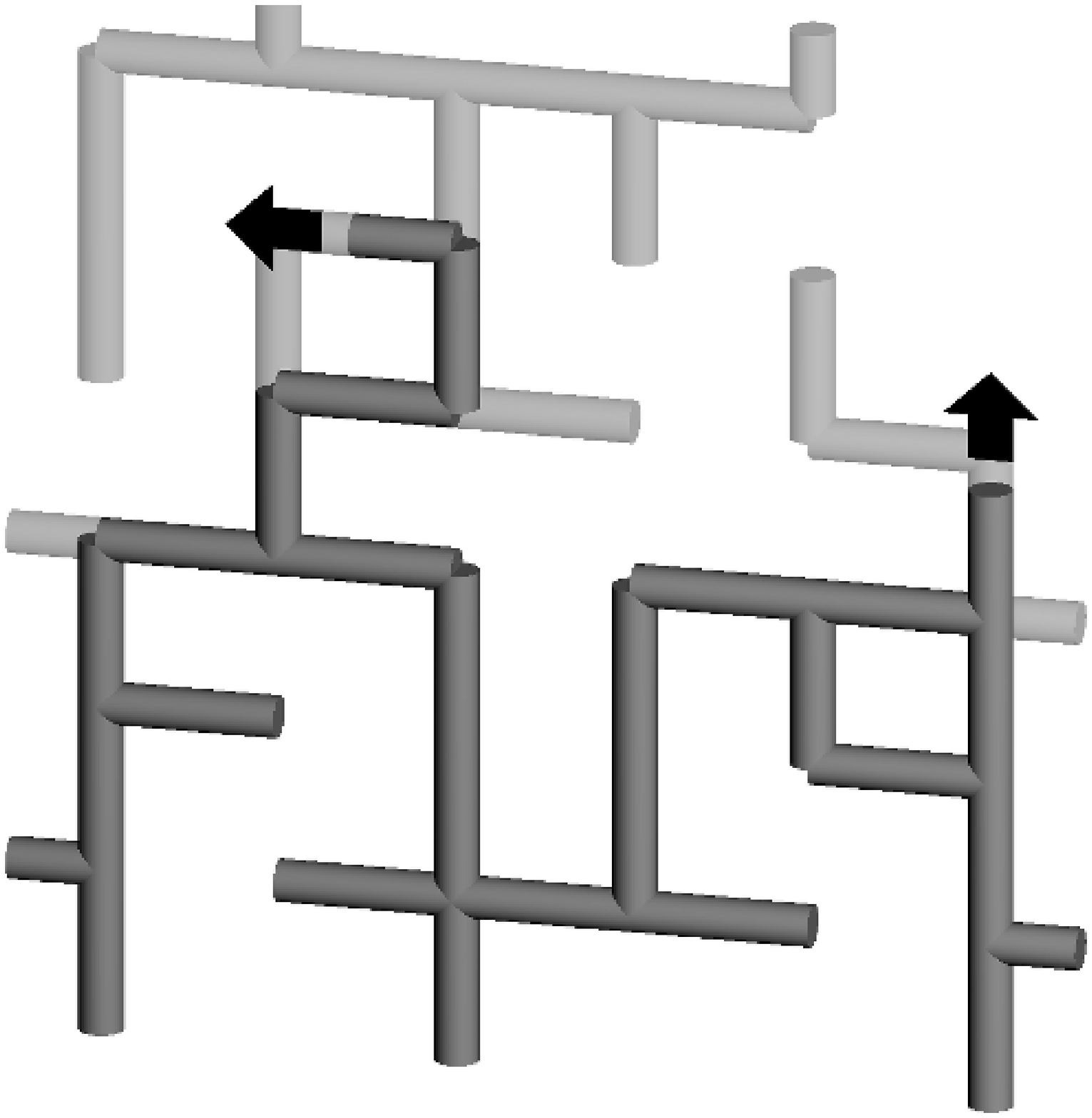}
\includegraphics[width=0.49\columnwidth]{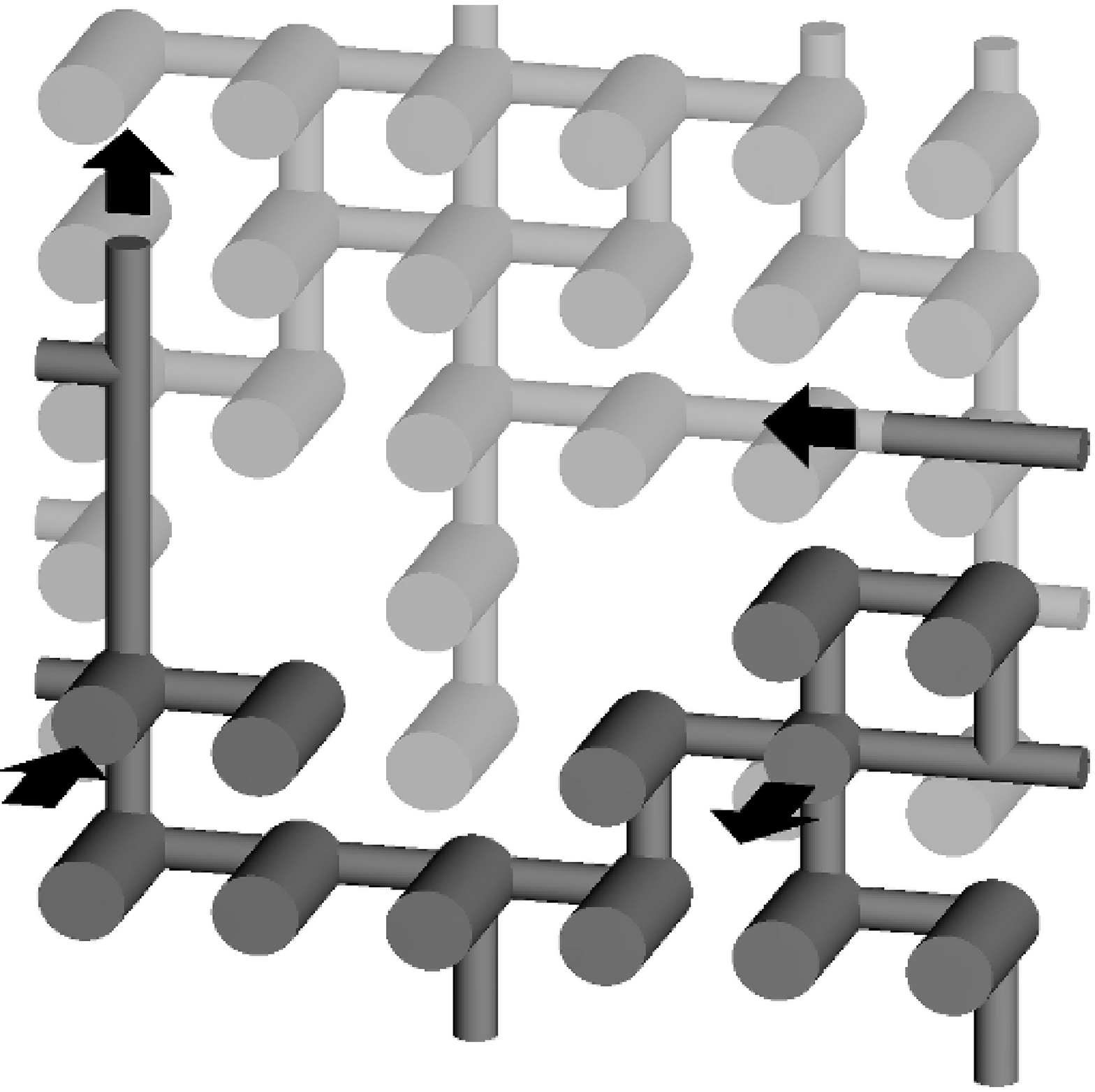}
\caption{
\label{fig:snapshot} 
Snapshots of fluid invasion for networks with low (a) and high (b)
volume capacitance at in-plane pipe population $p_0=0.53$. Wetted
(dry) regions are shaded in dark (light) gray. Radii are rescaled by a
factor 0.5 for clarity. Only connected pipes are shown. Arrows
indicate instantaneous directions of menicus movements. Dewetting is
forbidden in (a). In (b), one menicus recedes due to suction by other
invaded pipes with stronger capillary forces.  }
\end{figure}

Our models are based on a simple pipe network in Ref.
\cite{Lam} simulating the wetting of two-dimensional porous media with the
bottom immersed into a liquid reservoir. We adopt a square lattice
with spacing $a$ and periodic boundary conditions in the lateral
direction. 
% QB3
We define a in-plane bond population $p_0$ which is the probability
that a bond on the square lattice is occupied by a cylindrical pipe of
radius $r=0.25a$. A fraction $1-p_0$ of the bonds are left
vacant. 
%(See Fig. \ref{fig:snapshot}.)
%
Given a pressure gradient $\nabla P$ along a pipe, the liquid flows
according to Poiseuille's law for viscous flow with a flux ${\pi
r^4}\nabla P/{8\mu}$ where $\mu$ is the viscosity
\cite{Sahimi}. The atmospheric pressure is assumed to be zero without
loss of generality. Nodes at the bottom row are directly connected to
the liquid source and are also maintained at zero pressure. The pipes
are simplified models of complicated fluid channels. Without strictly
following the properties of realistic pipes, we assume that the
capillary pressure for an advancing meniscus inside a pipe follows an
uniform random distribution in the interval $[0,\Gamma_0]$ with $\Gamma_0$
being a constant. The liquid pressure behind a meniscus hence varies
from $-\Gamma_0$ to 0. All nodes are volumeless and gravity is
neglected.  Air flow as well as trapping are not considered.  A simple
system of units in which $a=\Gamma_0=\mu=1$ is assumed.

Two variants of the model are considered. They differ in the volume
capacitance defined as the volume of liquid extracted from the porous
medium locally per unit decrease in pressure
\cite{Furuberg}. We first define a low capacitance
network (Fig. \ref{fig:snapshot}(a)). In this case, we forbid any
receding meniscus by assuming an ideally hysteric capillary pressure
withstanding any tendency of dewetting. The liquid surface is rigid
and the volume capacitance is indeed zero. We also consider a high
capacitance network (Fig. \ref{fig:snapshot}(b)) in which receding
menisci are allowed and are non-hysteric. Furthermore, an extra
dangling pipe perpendicular to the lattice is connected to every
node. They all have length $a$ and radius $r=0.5a$. The capillary
pressure takes a random value uniformly distributed in $[0,
0.3\Gamma_0]$.
% QB3
This enhance the volume capacitance at pressure around $-0.3\Gamma_0$
to $0$ which is the relevant range since it coincides with typical
local pressures measured in wetted regions in our simulations. An
increase (decrease) in the local pressure within this range activates
the filling (draining) of many of these pipes resulting in the strong
capacitance enhancement. 
%The resulting pressure stabilization effects
%will further be discussed later.
%
We have checked that further increasing the capacitance by adding even
more dangling pipes or increasing their radii give similar
results. For simplicity, no new menisci can be generated by breaking
any continuous column of liquid during dewetting.

We simulate fluid invasion using Euler's method. For both models, the
pressure field is first computed at the beginning of each
iteration. Specifically, Poiseuille's law and fluid conservation at all
nodes leads to Kirchoff's equations coupling the pressure at
neighboring nodes. We also apply the boundary conditions that the
pressure at moving and stationary menisci follows from the capillary
pressure and the no flow condition respectively. The Kirchoff's
equations are then solved using successive over-relaxation method.
% QB2
For the low capacitance model with ideally hysteric capillarity,
meniscus in a partially filled pipe may start or stop advancement
depending on the calculated pressure. This alters the boundary
conditions and the whole calculation is repeated until the states of
motion of all menisci are consistent with the pressure field. For both
models, we speed up the calculation by directly computing only the
pressure at the percolation backbone.
We require that the liquid
influx from the source agrees with the total out-flux at the moving
menisci within 1\%. From the pressure, the velocities of all moving
menisci are found. Their positions are then advanced over a short
period in which all displacements are less than $0.1a$.

\begin{figure}
\includegraphics[width=\columnwidth]{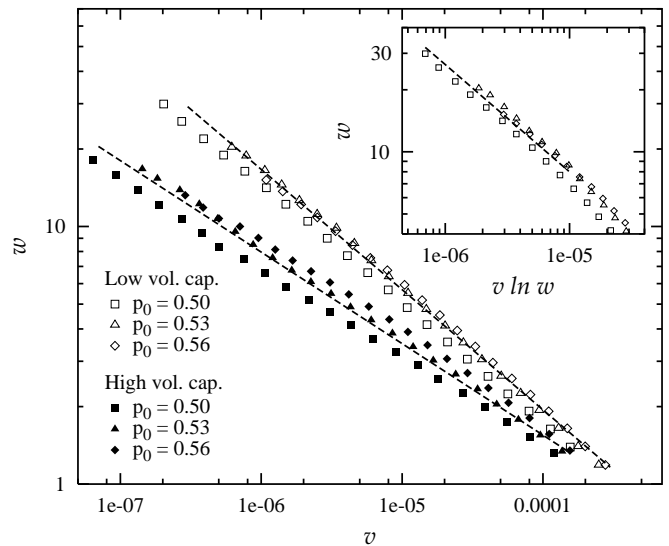}
\caption{
\label{fig:wv} 
Log-log plot of the invasion front width $w$ against velocity
$v$. The slopes of the dashed lines are -0.47 and -0.36
respectively. 
% QB4
Inset: log-log plot of $w$ against $v \ln w$. The dashed
line has a slope -0.53.
}

\end{figure}

Figure \ref{fig:wv} plots the front width $w$ against its velocity $v$
for both models at pipe population $p_0=0.50, 0.53$ and 0.56. The
lateral width of the lattice used is $L=1000a$ and all results have
been averaged over 60 independent runs. Results are extracted from 
surface height profiles $h(x)$ measured from the liquid source which
mark the highest invaded node at lateral coordinate $x$. The time
derivative of the spatial and ensemble average $\bar{h}$ gives $v$
while the r.m.s. fluctuation gives $w$.  The data fall nicely into
straight lines in the log-log plot and verify Eq. (\ref{wv}). Contrary
to conventional belief, there are two distinct sets of slopes
averaging to $\kappa = 0.47(2)$ and 0.36(2) for the low and high
volume capacitance networks respectively.

\begin{figure}
\includegraphics[width=8.5cm]{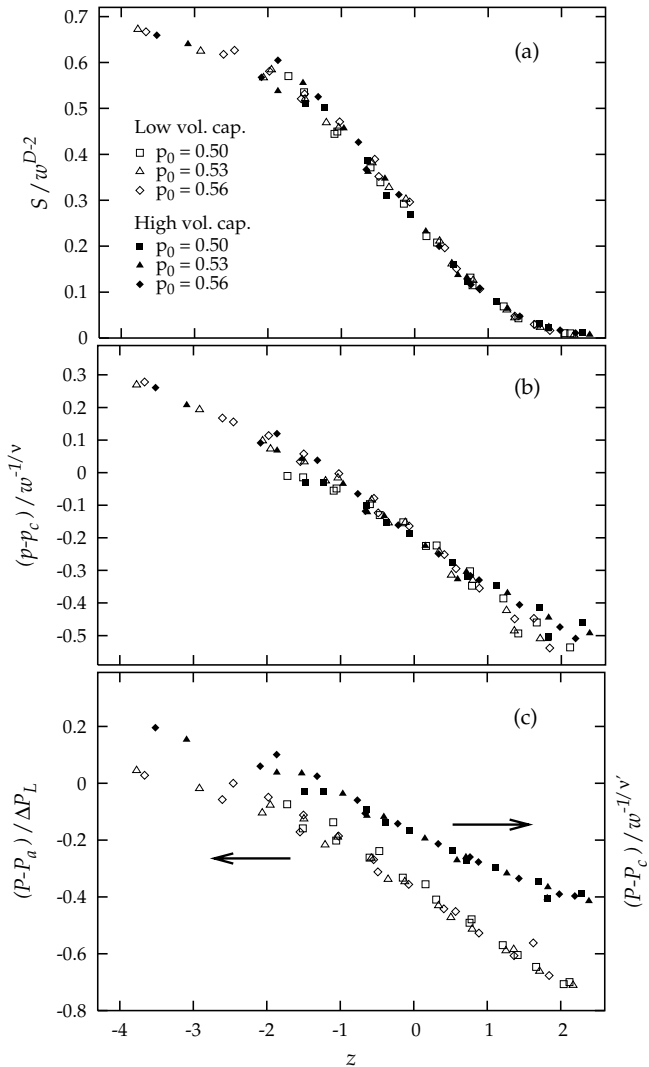}
\caption{
\label{fig:pden}
Rescaled plots of local saturation $S$ (a), percolation probability
$p$ (b), and pressure $P$ (c) against rescaled coordinate
$z=(y-\bar{h})/w$. Note that $P$ requires different rescaling forms
for the low and high volume capacitance cases.  }
\end{figure}

%\pagebreak

To attain better insights into the problem, we now examine closely the
geometry of the invasion fronts. Fronts at various locations and
widths will be compared. We hence define a rescaled distance $z$ from the
center of a front by 
%\begin{equation}
$z=(y-\bar{h})/w$
%z=\frac{y-\bar{h}}{w}
%\end{equation}
where $y$ is the vertical coordinate.  An important quantity is the
local saturation $S(y)$ defined as the fraction of {\it in-plane}
pipes which are completely invaded at height $y$. Figure
\ref{fig:pden}(a) plots $S(y)/w^{D-2}$ against $z$ using the same data
as in Fig. \ref{fig:snapshot} where $D=91/48$ is the fractal dimension
\cite{Stauffer}. Values recorded near the beginning
and the end of each set of simulations are plotted. Specifically, time $t$
goes from $1.8\times 10^5$ to $2.5
\times 10^8$ corresponding to $w$ in between 3.1 and 32. All 12 curves
collapse nicely into a single curve implying the scaling form
\begin{equation}
\label{S}
S(y) = w^{D-2} f_S (z)
\end{equation}
where $f_S$ is a rescaled function. This verifies the percolation
geometry of the fronts with $w$ being the correlation length
\cite{Stauffer}. Similarly, we define the local percolation
probability $p(y)$ as the probability that an in-plane pipe connected
to an invaded node is completely filled.  Figure \ref{fig:pden}(b)
plots $(p(y)-p_c)/w^{-1/\nu}$ against $z$ where $\nu=4/3$ and
$p_c=1/2$ are respectively the correlation length exponent and the
percolation threshold. Good data collapse is again observed supporting
\begin{equation}
\label{scale_p}
p(y)-p_c=w^{-1/\nu} f_p (z).
\end{equation}

\begin{figure}
\includegraphics[width=\columnwidth]{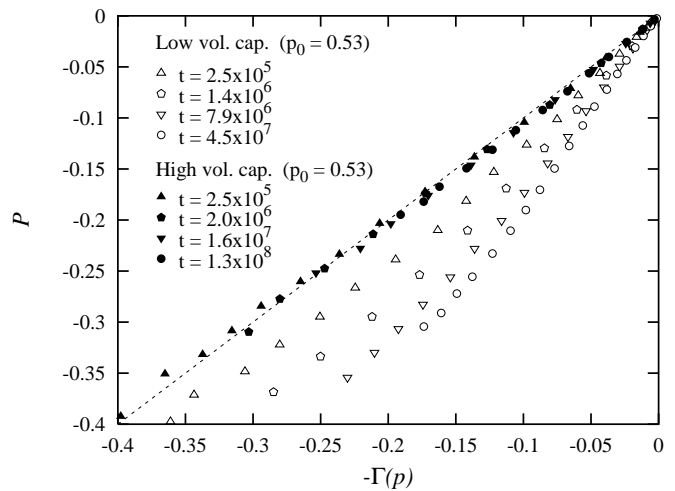}
\caption{
\label{fig:P} 
Plot of the average pressure $P$ against the instantaneous maximum
pressure $-\Gamma(p)$ expected from the percolation probability $p$.
}
\end{figure}

Data collapses for pressure in Fig. \ref{fig:pden}(c) are most
interesting and will be explained after examining the relation
between the local pressure and the local percolation
probability. During invasion, the percolation probability at any point
increases as pipes with ever lower capillary pressure are
filled. Given a local percolation probability $p$, pipes with
capillary pressure as low as $\Gamma(p) = ( 1 - p/p_0) \Gamma_0$ are
expected to be filled. There is hence at least one instance when the
pressure climbs up to instantaneous maximum value $-\Gamma(p)$. Figure
\ref{fig:P} compares the local pressure $P$ with $-\Gamma(p)$. Only
results for $p_0=0.53$ are shown but other values give similar
findings.

At high volume capacitance, Fig. \ref{fig:P} shows that the
pressure simply coincides with its upper bound, i.e.
\begin{equation}
\label{Pp}
P=-\Gamma(p)
\end{equation}
Letting $P_c=-\Gamma(p_c)$, it implies
\begin{equation}
\label{Pp2}
P(y) - P_c ~ \sim ~ p(y) - p_c
\end{equation}
which has been widely used in the literature
\cite{Xu,Wilkinson,Lenormand,Blunt,Aker} and is valid for gradient
percolation \cite{Hulin}. However, for viscous gradient percolation,
Eq. (\ref{Pp}) is indeed non-trivial and results from the strong
damping of pressure fluctuations. Local pressure depends on the
capillary pressure of the pipes under invasion. Pipes with capillary
pressure in the range $[\Gamma(p),\Gamma_0]$ are to be invaded. The
pressure is hence dominated by those with the smallest capillary
pressure $\Gamma(p)$ which takes the longest time to wet. Moreover, we
observe that when a pipe with a higher capillary pressure is invaded,
it quickly draws liquid from neighboring pipes, especially 
perpendicular ones with capillary pressure close to $\Gamma(p)$.
There is hence only a brief and highly localized pressure disturbance
which has little impact on the average pressure.  After completing the
invasion, these neighbors are slowly refilled directly from the liquid
reservoir and maintain the pressure roughly at $-\Gamma(p)$
again. This pressure stabilization mechanism leads to
Eq. (\ref{Pp}). Then, Eqs. (\ref{scale_p}) and (\ref{Pp2}) imply $P(y)
- P_c=w^{-1/\nu} f_P (z)$.  We therefore plot $(P(y)-P_c)/
w^{-1/\nu'}$ against $z$ for the high capacitance case and is shown in
Fig. \ref{fig:pden}(c) using filled symbols. We have taken $\nu'=1.5$
for the best data collapse which is in reasonable agreement with the
exact value $4/3$.
%QA1
The discrepancy inherits from the slight deviation from Eq. (\ref{Pp})
as observed in Fig. \ref{fig:P} and results from imcomplete damping of
the pressure fluctuations due to the finite volume capacitance of the
network.
The
pressure difference $\Delta P_H$ across the front is hence
\begin{equation}
\label{DP_high}
\Delta P_H \sim w^{-1/\nu}.
\end{equation}

At low capacitance, Fig. \ref{fig:P} implies $P < -\Gamma(p)$
instead. Pipes with capillary pressure larger than $\Gamma(p)$ can now
significantly pull down the pressure when invaded. A new analytic
description of the pressure field is now presented. We first
quantitatively define the front region by $z_a \le z \le z_b$. We take
$z_a=-2$ since it marks the bottom of the lattice for $p_0=0.50$ and
$z_b=2$ as it is where the saturation practically vanishes.  We denote
quantities evaluated at the boundaries by the subscripts $a$ and
$b$. For $z<z_a$ corresponding to the bulk region, good connectivity
suppresses pressure fluctuations and Eq. (\ref{Pp}) holds
approximately. In particular, the pressure at $z_a$ is $P_a = -
\Gamma(p_a)$. 
In contrast, the pressure fluctuates strongly at $z_b$ between
$-\Gamma_0$ and $-\Gamma(p_b)$ as pipes with capillary pressure in the
range $[\Gamma(p_b), \Gamma_0]$ are invaded. With dewetting
in other pipes forbidden, they draw liquid directly from the water
reservoir. This leads to strong
delocalized pressure fluctuations. The invasion of a pipe with
capillary pressure $P_{cap}$ takes a period $\tau \sim (
P_{cap}+P_a)^{-1}$. A characteristic pressure $P_b$ at $z_b$ hence follows
from the time averaged capillary pressure. Straight-forward
algebra gives $P_b = P_a - \Delta P_L$ where
\begin{equation}
\label{DP_low}
 \Delta P_L = [{\Gamma_0 - \Gamma(p_b)}]
/\ln \left[ \frac{\Gamma_0 + P_a}{\Gamma(p_b)+P_a} \right].
\end{equation}
Here $\Delta P_L$ represents a characteristic pressure difference
across the front. We thus suggest $P(y)-P_a = \Delta P_L f_P(z)$. It
is verified by the reasonable data collapse in Fig. \ref{fig:pden}(c) which plots
$(P(y)-P_a)/\Delta P_L$ against $z$ for the high capacitance case
using open symbols. We have used $p_{a,b} = p_c + w^{-1/\nu}
f_p(z_{a,b})$ from Eq. (\ref{scale_p}) and further assumed
$f_p(z_a)=0.15$ and $f_p(z_b)=-0.5$ which are consistent with values
read directly from Fig. \ref{fig:pden}(b).

Now, we can calculate the exponent $\kappa$ defined in
Eq. (\ref{wv}). The permeability of the front is $k
\sim w^{-\mu/\nu-1}$ where $\mu$ is the conductivity exponent
\cite{Stauffer}. The liquid flux across the front is then $k\Delta
P_{H,L}$ for high and low capacitance respectively. However, it should
equal $vS$ as well. Using also Eq. (\ref{S}), we obtain $v \sim w ^
{\mu/\nu + D -1} \Delta P_{L,H}$.  For the high capacitance model
obeying Eq. (\ref{DP_high}), we arrive at Eq. (\ref{wv}) with
\begin{equation}
\label{kappa-h}
\kappa = \nu / (1+\mu + \nu(D-1)) \simeq 0.38 
\end{equation}
derived previously in Refs. \cite{Xu,Wilkinson,Aker}.  Our network
simulations in the high capacitance regime lead to $\kappa\simeq
0.36(2)$ giving a direct numerical support to Eq. (\ref{kappa-h}).

For the low capacitance case, we apply Eq. (\ref{DP_low}) which
simplifies at large $w$ to $\Delta P_L \sim (\ln w)^{-1}$. Therefore,
Eq. (\ref{wv}) is replaced by
\begin{equation}
v \sim \frac{w^{-1/\kappa}}{ \ln w} 
\end{equation}
with a {\it new} exponent
%We get $v \sim  ( \ln w) ^{-1} w^{-1/\kappa}$ with
%\sim  (v\ln \frac{1}{v})^{-\kappa} $ with
\begin{equation}
\label{kappa-l}
\kappa = \nu / (\mu + \nu(D-1)) \simeq 0.53 
\end{equation}
The scaling thus admits a logarithmic correction.
% QB4
This is verified by the log-log plot of $w$ against $v\ln w$ in the
inset of Fig. \ref{fig:wv} which gives straight lines with the
expected slope -0.53(2) for $w \agt 8$.
A naive measurement of $\kappa$ from a log-log plot of $w$ against $v$
disregarding the correction should lead to an effective exponent
$\kappa_{e}$ in between $0.38$ and $0.53$. Using Eq. (\ref{DP_low}),
we obtain $\kappa_{e}\simeq 0.46$ in good agreement with the value
$0.47(2)$ directly measured from our simulations and 0.48 from
imbibition experiments in Refs.
\cite{Horvath,Rubio}.

We have considered extreme values of the volume capacitance. In
general, networks with significant pressure fluctuations extending
beyond or well-within the invasion front are at the low or high
capacitance limit respectively. Networks with intermediate capacitance
are therefore expected to crossover to the high capacitance regime at
the large front width limit. However, from experiments, pressure
fluctuations in the form of Haines jumps propagate far beyond the
invasion front
\cite{Furuberg}. It is also known that wetting and drainage in porous
media are highly hysteric. These further establish that the new
low-capacitance condition is the correct experimentally relevant
regime. In addition, similar to the original model in Ref. \cite{Lam},
our low capacitance model can also reproduce the rich dynamical
features of the invasion front in Ref. \cite{Horvath} with reasonable
accuracy at $p_0=0.53$ and will be reported elsewhere.
%Our results should apply to both drainage and imbibition when
%complications such as film flow and snap-off are negligible
%\cite{Blunt}.  
%It will be interesting to generalize our results to more
%general situations including a finite viscosity of the displaced phase
%as well as trapping \cite{Xu,Aker,Meheust,Ferer}.  
%
%Models based on
%interface pinning have been suggested to describe experiments
%on imbibition \cite{Dube}. However, anisotropic evolution rules are
%often assumed but can be inconsistent with experiments
%\cite{Albert}.  
%
%Moreover, many previous studies focus only on a
%roughness exponent which in many cases is not accurately reproducible
%experimentally. 
%In contrast, network models are well established for
%flow in porous media \cite{Sahimi} and has successfully reproduce many
%observations in experiments
%\cite{Lam}. The theoretical understanding attained here should motivate
%more critical cross checking between network simulations and other
%approaches.

In conclusion, we have derived a new effective scaling relation based
on viscous gradient percolation for fluid invasion fronts in porous
media with low volume capacitance. This regime is characterized by
realistic features including highly hysteric capillary flow and
long-range propagation of pressure disturbances. The scaling is
supported by large scale network simulations and agrees with previous
experiments. We also show numerically that a well-known scaling theory
applies only at high volume capacitance which is not supported
experimentally.

We thank V.K. Horv\'ath, L.M. Sander, J.Q. You and F.G. Shin for
helpful discussions. This work was supported by HK RGC Grant
No. PolyU-5191/99P.

\end{document}